\documentclass[
superscriptaddress,
twocolumn,
amsmath,amssymb,
aps,
prl,
]{revtex4-1}
\usepackage{color}
\usepackage{gensymb}
\usepackage{graphicx}
\usepackage{dcolumn}
\usepackage{bm}
\usepackage{hyperref}
\usepackage{epstopdf}
\usepackage{float}
\usepackage{listings}
\usepackage{times,mathptm,bm}
\usepackage{xcolor}
\usepackage[export]{adjustbox}

\input ulem.sty\relax
\catcode`\@=11 %

\font\uwavefont=lasyb10 scaled 652

\def\uwave{%
  \bgroup
    \markoverwith{%
      \lower3.5\p@\hbox{\uwavefont\char58}%
    }%
  \ULon
}

\begin{document}

\title{Phase Separation by Entanglement of Active Polymer-like Worms}

\author{A. Deblais}
\affiliation{Van der Waals-Zeeman Institute, Institute of Physics, University of Amsterdam, 1098XH Amsterdam, The Netherlands.}
\author{A. C. Maggs}
\affiliation{UMR Gulliver 7083 CNRS, ESPCI, PSL Research University, 10 rue Vauquelin, 75005 Paris, France.}
\author{D. Bonn}
\affiliation{Van der Waals-Zeeman Institute, Institute of Physics, University of Amsterdam, 1098XH Amsterdam, The Netherlands.}
\author{S. Woutersen}
\affiliation{Van 't Hoff Institute for Molecular Sciences, University of Amsterdam, Science Park 904, 1098XH Amsterdam, The Netherlands.}

\date{\today}

\begin{abstract}
We investigate the aggregation and phase separation of thin, living {\it T.~Tubifex} worms that behave as active polymers. Randomly dispersed active worms spontaneously aggregate to form compact, highly entangled blobs, a process similar to polymer phase separation, and for which we observe power-law growth kinetics. We find that the phase separation of active polymer-like worms does not occur through Ostwald ripening, but through active motion and coalescence of the phase domains. Interestingly, the growth mechanism differs from conventional growth by droplet coalescence: the diffusion constant characterizing the random motion of a worm blob is independent of its size, a phenomenon that can be explained from the fact that the active random motion arises only from the worms at the surface of the blob. This leads to a fundamentally different phase-separation mechanism, that may be unique to active polymers.
\end{abstract}

\pacs{Valid PACS appear here}
\maketitle

Driven colloidal particles \cite{palacci13,Geyer2019, Linden2019}, self-propelled bots \cite{Briand2016, Deblais2018}, cells~\cite{Schwarz-Lineka2012, Duclos2018}, animals~\cite{Tennenbaum2015,sugi19} and humans~\cite{Bain2019} belong to the field of active matter: interacting agents that extract energy from the environment to produce sustained motion or mechanical stresses \cite{liverpool00,speck14,Linden2019}. Their collective behavior is fascinating, and the activity and interactions of the individual components give rise to highly non-trivial macroscopic phenomena \cite{Bechinger2012}. Here, we investigate the spontaneous aggregation and  phase separation of active-polymer-like living worms. At first sight, the phase separation of active polymers may seem similar to that of solutions \cite{Dobry1947, Tanaka2005} and crystallizing solids \cite{Olson1988}, for which detailed theories are available: in both cases, the aggregating particles move randomly and tend to stick together when they are in close proximity. As such, one might expect the phase separation of active particles to involve a mechanism similar to Ostwald ripening \cite{Voorhees1992}, where the aggregation is driven by the combined effect of diffusion of the aggregating particles and surface tension of the aggregates. However, recent work has shown that the phase separation of active particles can involve mechanisms that rely on the activity~\cite{Bechinger2012,Schwarz-Lineka2012,Thompson2011,Cates2015,gonnella15,Geyer2019,Linden2019}, and our results indicate that this also holds for the phase separation of active polymers.

\begin{figure}[b]
\begin{center}
\includegraphics[width=8.5cm]{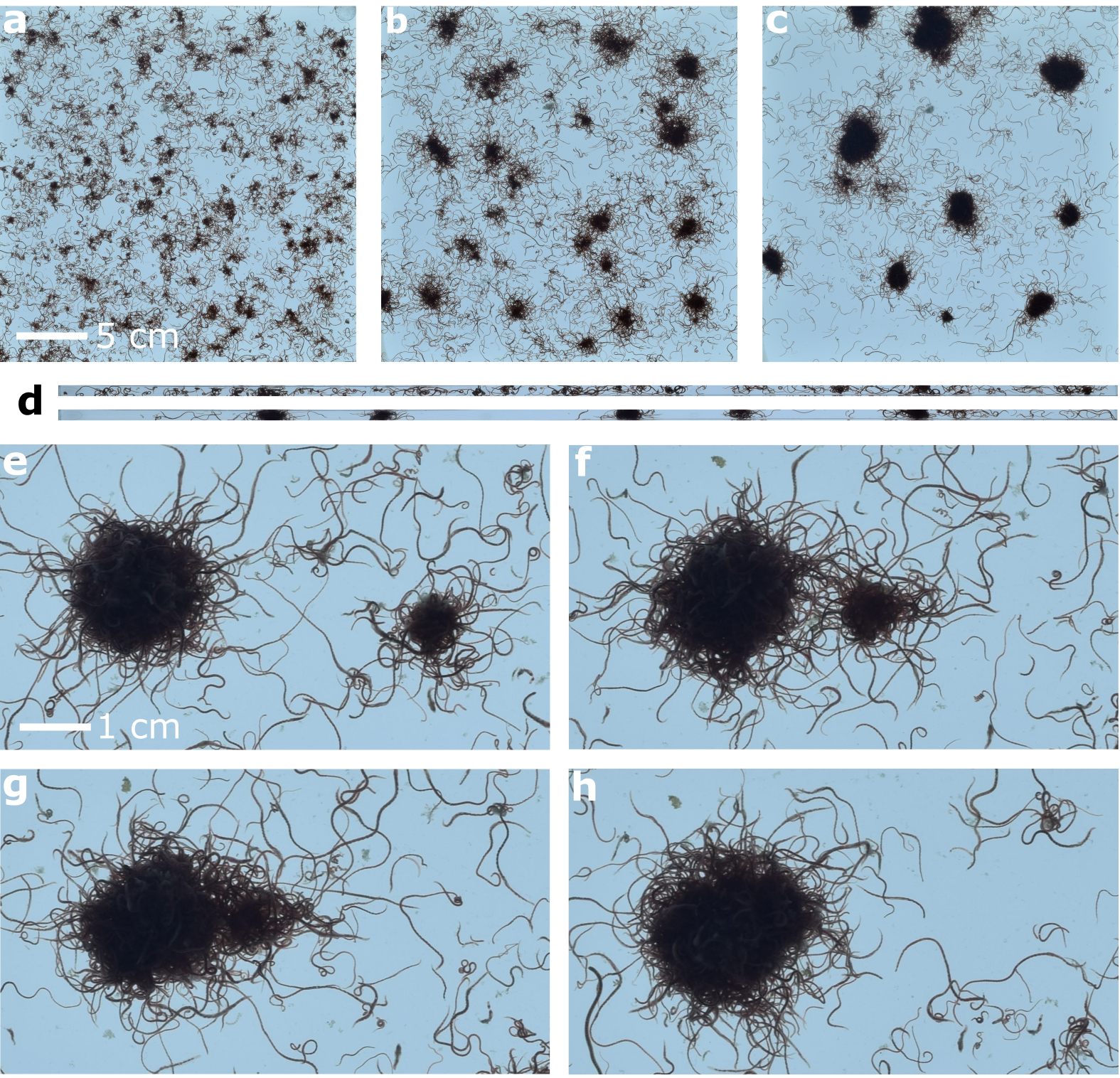}
\end{center}
\vspace{-2em}
\caption{\textbf{Aggregation and phase separation of \textit{T.~Tubifex}}. (\textbf{a--c})~ Snapshots of active-worm aggregation in a 25$\times$25$\times$2.5~cm volume at $t=0$~(a), 9.5~min~(b), and 60~min~(c). \textbf{d}~Snapshots from a 1D experiment in a square tube of 51$\times$1$\times$1$\times$1~cm. (\textbf{e--h})~Close ups from another experiment at $t=25$~min~(e), 28.5~min~(f), 29~min~(g), and 31.5~min~(h), showing the coalescence of two blobs. See \cite{supmat} for the videos.} 
\label{fig:photos}
\end{figure}

\begin{figure*}[t]
\begin{center}
\includegraphics[width=\linewidth]{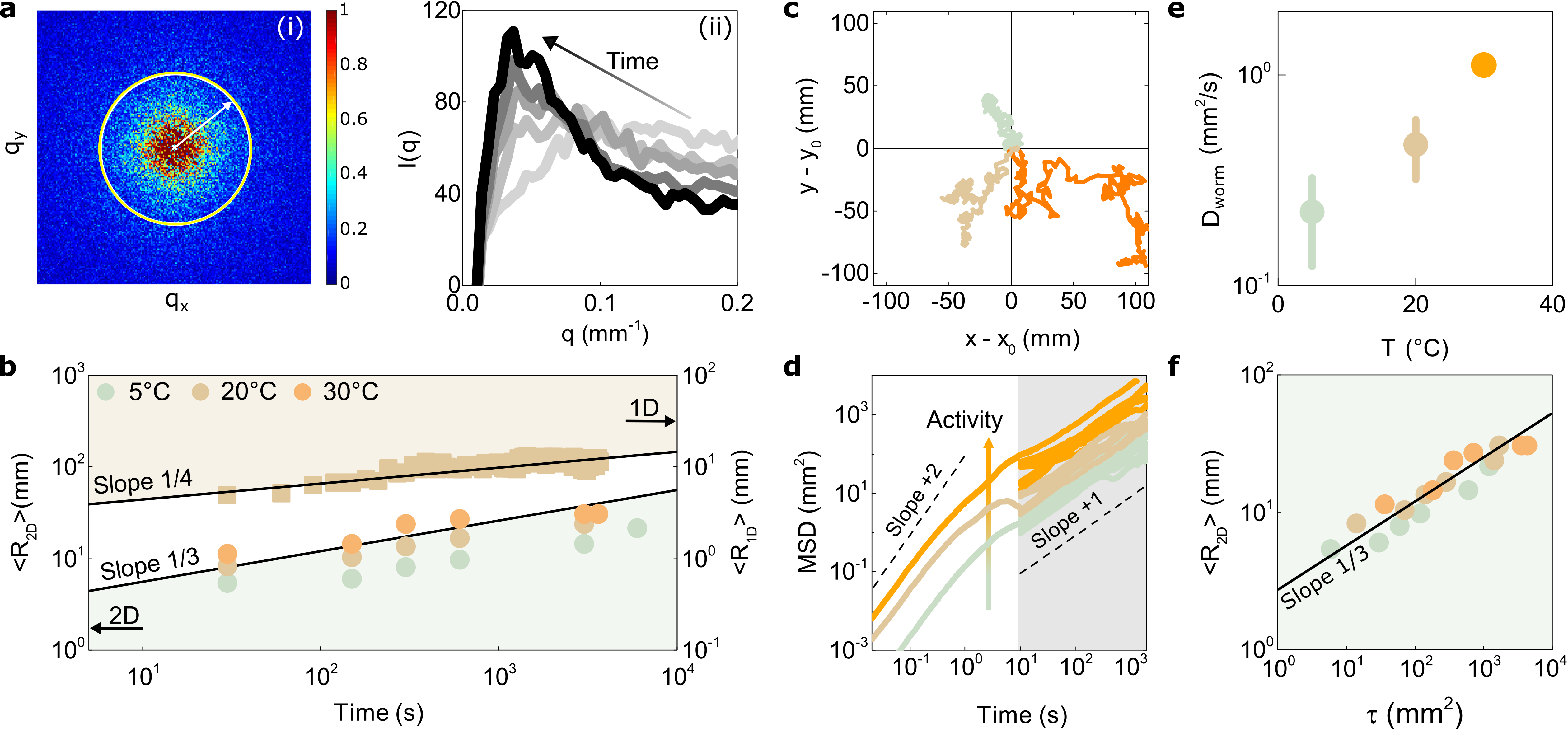}
\end{center}
\vspace{-2em}
\caption{\textbf{Phase-separation dynamics.} \textbf{a}~ (i) Modulus squared of the Fourier transform and (ii) Fourier power spectrum of the image of the spatial worm distribution at different times. The $q$ value at which the intensity is maximal shifts to lower values with increasing time (2D experiment, T=30$\degree$C). \textbf{b}~Average size  $\langle R_{nD} \rangle \sim q_{\rm max}^{-1}$ (left axis for 2D experiments and right axis for 1D experiments) as a function of time, showing approximate power-law behavior with a power of 1/3 for the 2D experiment at all temperatures (circles) and 1/4 for the 1D experiment at ambient temperature (squares). The solid lines are guides to the eye. \textbf{c}~Trajectories of a single worm time at a controlled temperature $T$. Each solid trace of a different colour represents the centre-of-mass trajectory of one-hour duration for the same worm at a different temperature ($T=5,20,30 \degree$C; color coding as in \textbf{b}). The origin of all trajectories is set to $(x_{0},y_{0})=(0,0)$. \textbf{d}~Mean square displacement (MSD) as a function of time of ten worms in water at $T=5,20,30 \degree$C (same color coding as in \textbf{b}). The dashed line of slope 1 in the shaded area shows the expected scaling for Brownian motion. The dashed line in the non-shaded area has slope 2. \textbf{e}~Diffusion coefficient $D_{\rm worm}$ of a worm extracted from the MSD(t) as a function of temperature. Each measurement is an average over 10 trajectories of the same worms. \textbf{f}~ Average size of all the 2D experiments at the different temperatures collapsed onto a single master curve by rescaling the time $\tau = D_{\rm worm} t$.}
  \label{fig:fourier}
\end{figure*}

The {\it T.~Tubifex} worms that we investigate are active swimmers, and approximately 300\ $\mu$m thick and 10--40~mm long~\cite{supmat}. The thermal random motion of the worms (estimated from the Stokes-Einstein equation) is negligible compared to their active motion, so they constitute a good model system for active polymers \cite{winkler17,martin18,Bianco2018,martin19,Mokthari2019,deblais_preprint_2019}. When randomly distributed over a volume of water, the worms aggregate spontaneously (Fig.\ 1) into highly entangled `blobs'. 
In the remainder we use the term `phase separation' for the final state in which there are macroscopic domains (i.e.\ containing large numbers of worms) of high worm density in a space with essentially zero worm density, and `aggregation' to refer to the growth of these domains from individual worms.
The biological function of this aggregation is to minimize exposure to dissolved oxygen, high levels of which are poisonous to {\it Tubifex}~\cite{walker1971}. The worms  
cling together by entanglement, which is aided by small bristles on their bodies~\cite{brinkhurst63}. The aggregation is irreversible on the time scale of the experiment and the worms are submerged in water, which makes the aggregation different from motility-induced phase separation~\cite{gonnella15} and from the recently studied aggregation of {\it C.\ elegans}, which forms dynamical bundle-shaped networks kept together mainly by surface tension of the water around the worms~\cite{sugi19}. We can analyze the aggregation of {\it T.\ Tubifex} with relatively simple methods, which makes these living worms an excellent system to investigate the phase separation of active polymers.

In the experiments, we disperse a specific number of worms in a thermostated water volume and observe their aggregation in real time by recording videos~\cite{supmat}. We investigate two geometries: in the simplest, the worms are dispersed in a 25$\times$25$\times$2.5~cm volume of water. In this geometry worm motion is effectively two-dimensional (2D), since the worms are denser than water and therefore always located at or close to the bottom of the water volume (Fig.\ S3). In other experiments the water volume is a 51$\times$1$\times$1~cm channel in which the worms are confined in an effectively one-dimensional space (1D). Figures~\ref{fig:photos}a--c show snapshots from a typical 2D experiment; Fig.~\ref{fig:photos}d from a typical 1D experiment~\cite{supmat}. In both geometries, as time progresses the worms form ever larger aggregates until after about 1 hour all worms are concentrated into a few large aggregates, which have the shape of a slightly flattened sphere~\cite{supmat}, a compromise between minimum exposed surface and minimum gravitational energy. In the following, we refer to these worm aggregates as `blobs'. 

The aggregation of the active worms seems similar to that generally observed for polymer phase separation. Hence, one might expect that the aggregation occurs through Ostwald ripening, in which larger aggregates grow at the cost of smaller ones~\cite{jonesbook}. This mechanism is driven by the reduction of the total surface tension with increasing average blob size, and we do in fact measure a finite surface tension for the worm blobs~\cite{supmat}. However, closer inspection (Figs.~\ref{fig:photos}~e--h) shows that blobs of all sizes are growing, and that the growth does not occur by Ostwald ripening, but rather by the merging of smaller aggregates into larger ones.
Such growth by coalescence of diffusing droplets has been investigated previously in the context of the growth of vapor-deposited thin films and of droplets on a surface~\cite{family88,steyer91,meakin92}.

To quantify our observations, we characterize the average blob size $\langle R  \rangle$ as a function of time by taking the 2D Fourier transform of the images and determining the wave-vector magnitude $q_{\rm max}$ at which the spectrum reaches its maximum intensity (Fig.~\ref{fig:fourier}a). 
Since the worm density in a blob is independent of its size~\cite{supmat}, the size of a blob gives direct information on the number of constituent worms.
With increasing time,  $q_{\rm max}^{-1}$ shifts to lower values, corresponding to an increasing average blob size $\langle R \rangle$. Figure~\ref{fig:fourier}b shows the average blob size $q_{\rm max}^{-1}$ determined in this way as a function of time, at several temperatures. The growth exhibits power law behavior with $\langle R_{2D} \rangle \sim t^{1/3}$ in the 2D experiments, and $\langle R_{1D} \rangle \sim t^{1/4}$ in the 1D experiments, independently of the initial worm concentration (Fig.\ S6). In addition, we observe in the 2D experiments that the power-law growth behavior is independent of temperature. 

We also investigate the  motion of individual worms by recording image sequences of single, isolated worms at different temperatures~\cite{supmat}. Figure \ref{fig:fourier}c shows an example of the centre-of-mass trajectory of an individual worm at various temperatures. The motion is a random walk with an effective diffusion constant that increases with temperature, as is confirmed by extracting the mean square displacement from which we retrieve the diffusion coefficient~(Fig.\ \ref{fig:fourier}e)~\cite{supmat}. By rescaling the time axis $\tau \propto D_{\rm worm} t$ all the growth curves collapse onto a single master curve $\propto\tau^{1/3}$ (Fig.~\ref{fig:fourier}f), confirming that the aggregation kinetics is determined by the random motion of the the worms. Unlike {\it C.\ elegans} which forms parallel bundles~\cite{sugi19}, the worms are not aligned but highly entangled in the blobs (Fig.\ S4c), and the probability of two worms coalescing is independent of the angle between their velocities during collision~\cite{supmat}, probably because their motion is wriggling rather than slithering.

To shed more light on the observed power-law growth kinetics, we perform computer simulations of 2D growth by coalescence of diffusing droplets, using an approach similar to that of Ref.~\citenum{steyer91} (see \cite{supmat} for details). We assume that the blobs are spherical and move randomly in a 2D space, and that two blobs with radii $R_1$ and $R_2$ at positions ${\bf r}_1$ and ${\bf r}_2$ coalesce when $|{\bf r}_1-{\bf r}_2|<R_1+R_2$ to form a new blob with radius $R=\sqrt[3]{R_1^3+R_2^3}$ at position ${\bf r}=(R_1^3 {\bf r}_1+R_2^3{\bf r}_2)/(R_1^3+R_2^3)$. The simulation starts with a random distribution of mono-disperse spheres with radius 1 which represent the individual worms, in a 2D space of size 200$\times$200. At every time step, each droplet is moved in a random direction by a distance equal to the diffusion constant $D_{\rm blob}$. 

\begin{figure}[t]
\begin{center}
\includegraphics[width=8.5cm]{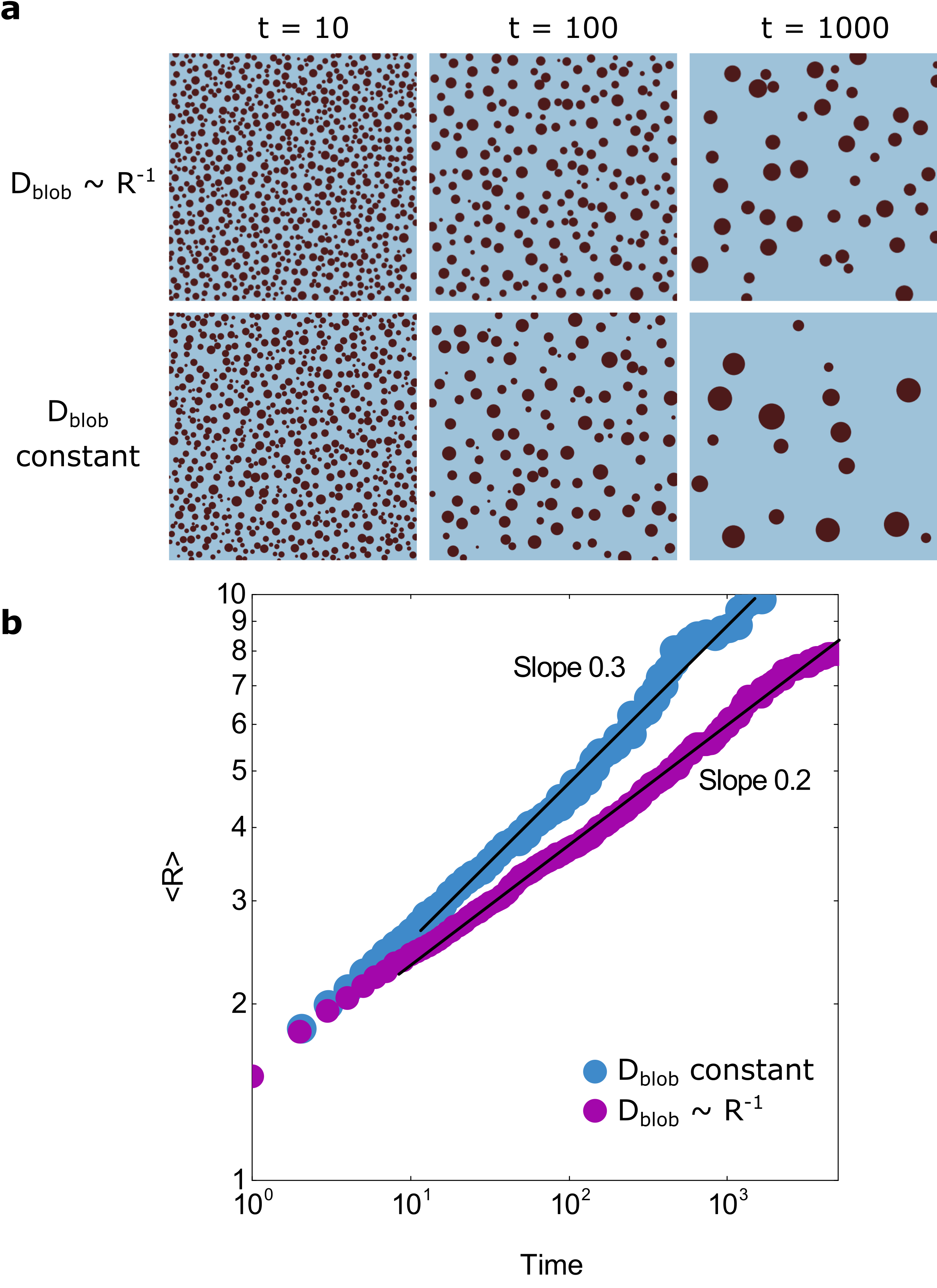}
\end{center}
\vspace{-2em}
\caption{\textbf{Simulation of phase-separation dynamics.} \textbf{a}~Snapshots of simulated blob growth by coalescence of randomly moving spherical blobs, with blob diffusion constant $D_{\rm blob}$ inversely proportional to the blob radius (top), and independent of blob radius (bottom). \textbf{b}~Average blob radius $\langle R \rangle$ as a function of time obtained from the simulations. In all cases the growth follows a power law (indicated by lines), with exponents of 0.2 and 0.3 respectively.}
  \label{fig:sim}
\end{figure}

Fig.\ \ref{fig:sim}a shows snapshots from two simulations with two different $D_{\rm blob}(R)$ functional dependencies, and Fig.\ \ref{fig:sim}b the time-dependent mass-weighted average size $\langle R \rangle = \sum_i R_i^4 /\sum_i R_i^3$, where both summations run over all the spheres present in the volume (similar results are obtained when using the number-weighted average, see Fig.\ S4). If we assume that the effective diffusion constant $D_{\rm blob}$ of a randomly moving blob depends on its radius $R$ as $D_{\rm blob}=R^{-1}$, as is the case for blobs undergoing Brownian random motion (Stokes-Einstein equation)~\cite{jonesbook}, we obtain power-law growth of the average blob size with an exponent of $\sim$0.20 (purple circles in Fig.\ \ref{fig:sim}b). 
This exponent is much smaller the experimentally observed value of $\sim$0.3 (Fig.~\ref{fig:fourier}f). It may be noted that the exponent obtained from the simulation is  different from that in Ref.~\cite{steyer91} because in our case we have conservation of total mass, and so depletion in the space between the blobs. 

\begin{figure}[t]
\begin{center}
\includegraphics[width=7.cm]{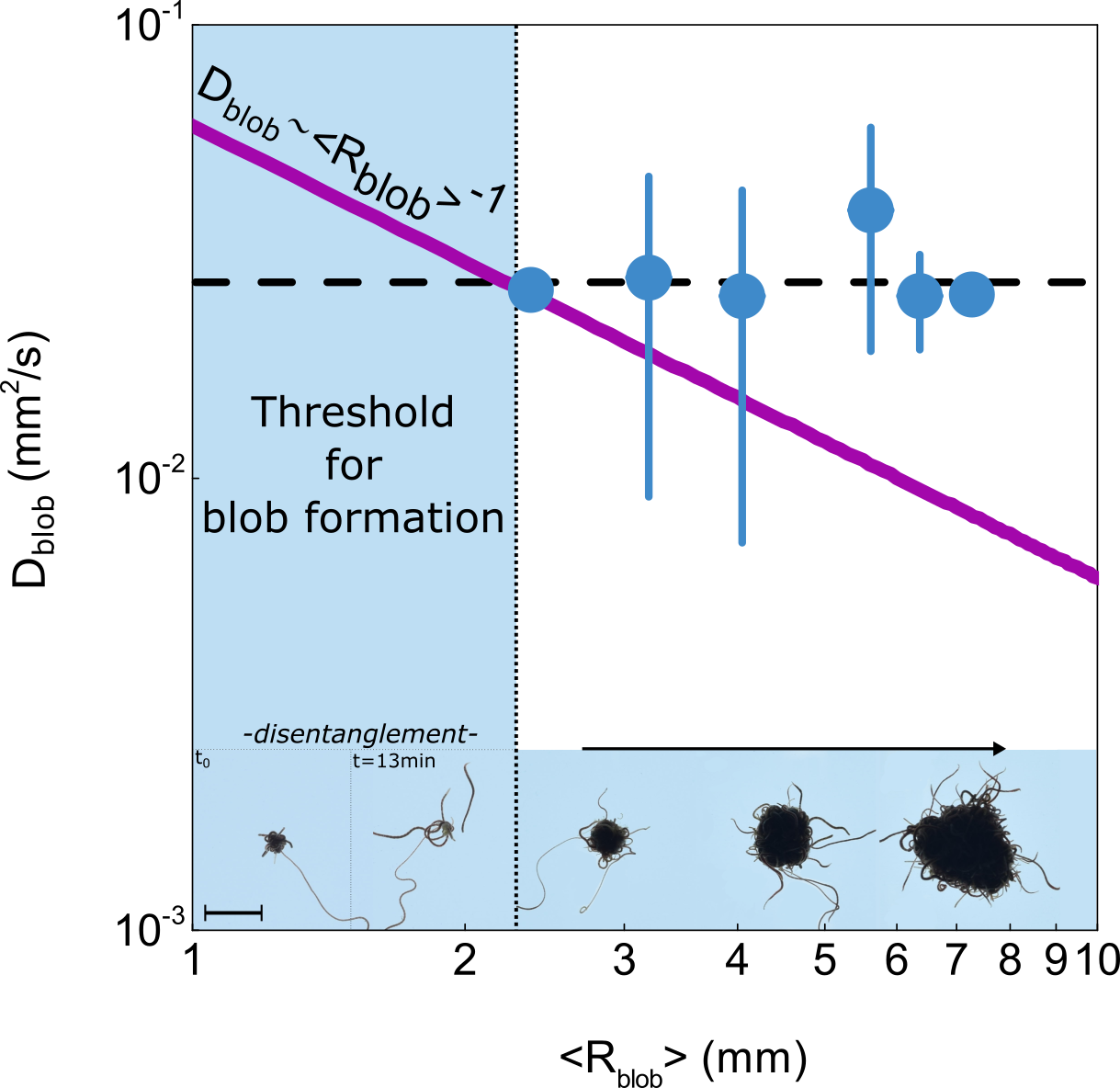}
\end{center}
\vspace{-2em}
\caption{\textbf{Blob diffusion.} Effective diffusion constant as a function of average blob size $\langle R_{\rm blob} \rangle$ at T=20$\degree C$ as determined from the slopes of experimental MSDs~\cite{supmat}. The error bars are mostly due to sample-to-sample variability. The purple line shows the expected scaling for particles undergoing Brownian random motion ($D_{blob} \sim {\langle R_{\rm blob} \rangle}^{-1}$). The experimental data (blue symbols) indicate a diffusion constant independent of blob size (dotted line). Lower insets: photographs of blobs of different sizes from which we measured the diffusion coefficient.}
  \label{fig:blobdiff}
\end{figure}

We believe that the discrepancy between the simulated and experimentally observed power-law exponents is due to the active motion of the worms. For a blob of active worms, the effective diffusion constant characterizing the random motion may not be inversely proportional to the blob size, as it is for a particle undergoing conventional, Brownian random motion~\cite{jonesbook}. To investigate this idea in more detail, we perform additional experiments in which we determine the effective diffusion constants of blobs with different sizes (Fig.~\ref{fig:blobdiff}). 

Interestingly, the effective diffusion constant does not depend on $R_{\rm blob}$ as $D_{\rm blob}\propto R_{\rm blob}^{-1}$, but appears to be independent of the radius of the blob. This means that the random motion of the worm blobs differs from conventional, Brownian random motion,
where $D_{\rm blob}=k_{\rm B}T/6\pi\eta R_{\rm blob}$ with $k_{\rm B}$ Boltzmann's constant and $\eta$ the viscosity of the surrounding liquid~\cite{jonesbook}. The difference can be explained by considering the origin of the random motion: Brownian random motion of a particle is caused by the thermal motion of the surrounding molecules, whereas the random motion of a worm blob is due to the active motion of the constituent worms, and this can lead to a different dependence of $D_{\rm blob}$ on $R_{\rm blob}$.

The observed size-independent blob-diffusion constant for active worms can be rationalized using a simple model. Let us assume that an individual worm $i$ at the bottom of a blob exerts a force $\pmb{f}_i(t)$ on the surface with rms magnitude $f_0=\sqrt{\langle |\pmb{f}_i(t)|^2\rangle}$, in a direction that varies randomly with a correlation time $\tau$, i.e.\ $\langle \pmb{f}_i(t)\cdot\pmb{f}_i(0)\rangle=f_0^2 e^{-t/\tau}$ (for simplicity we take $f_0$ and $\tau$ to be the same for all worms, but allowing a distribution of values does not change the main result). The total force exerted by $N$ worms is $\pmb{F}(t)=\sum_i \pmb{f}_i(t)$, and assuming zero cross correlation between the individual worm forces we have
$\langle|\pmb{F}(t)|^2\rangle=\sum_i\langle |\pmb{f}_i(t)|^2\rangle=Nf_0^2$, so the rms magnitude of the total force is $\sqrt{N}f_0$, and
$\langle\pmb{F}(t)\cdot\pmb{F}(0)\rangle=Nf_0^2e^{-t/\tau}$.
Since the worms inside the entangled blob are effectively immobilized, only the worms at the outer surface of a blob contribute force. The number of worms at the surface of a blob of radius $R_{\rm blob}$ is $N\sim R_{\rm blob}^2$, so the total random force exerted by these $N$ worms has a rms magnitude $\sqrt{N}f_0\sim Rf_0$, and a correlation time $\tau$. The drag force on the blob as a function of speed $v$ is given by Stokes' law, $F_{\rm drag}\sim vR_{\rm blob}$. To obtain the steady-state speed $v$ we equate the driving and drag forces, and obtain a velocity $v$ that is independent of the blob radius $R_{\rm blob}$. The random walk of a blob occurs by random steps (of duration $\sim\tau$) in which it has approximately the steady-state velocity $v$ (independent of size), and so the blob-diffusion constant is independent of blob size. It might be that worms at the bottom of a blob have slightly less activity than the ones on the sides and on the top, but this would not change the essentials of the scaling argument: the blob shape is to a good approximation independent of size (Fig.\ S5a), so the fraction of surface worms located at the bottom surface is approximately constant, and taking their potentially reduced activity into account would only give rise to a size-independent prefactor in the scaling expression for the total force exerted by the worms. 

Again simulating the blob growth, but now using a size-independent blob diffusion constant ($D_{\rm blob}=1$), we obtain a growth exponent of $\sim0.3$ (red points and line in Fig.~\ref{fig:sim}b), in good agreement with the experimentally observed value. Thus a size-independent diffusion constant for the random motion of the blobs explains the observed power-law growth of the worm blobs (Figs.~\ref{fig:fourier} and \ref{fig:sim}), at least in the 2D experiments.  
In the 1D experiments, the situation is complicated by the fact that blob motion slows down when the blob size becomes comparable to the channel width, possibly due to friction at the glass/worm surface. This makes quantitative analysis of the blob diffusion difficult. However, the fact that in this case the diffusion constant does decrease with $R_{\rm blob}$, and that at the same time the growth exponent is closer to that predicted by the conventional droplet-coalescence model does provide a qualitative confirmation of the above ideas.

The coalescence of blobs can be modeled by the Smoluchowski aggregation equation, but here we give a scaling argument for the dynamics.
We assume stochastic coalescence of the blobs and a size-independent blob-diffusion constant. We characterize the blob-size distribution by a single blob radius $R(t)$ that increases with time and start the system with a mass per unit area $\sigma$.  
For worms of unit mass density, the mass of a single blob is
$M_{blob} \sim R^3$.
The number density of blobs in the plane is $n(t)$ such that
$n(t) R^3 = \sigma$,
so the separation $a(t)$ between aggregates varies as 
$a(t) \sim 1/\sqrt{n(t)}=\sqrt{R^{3}(t)/\sigma}$.
The blobs diffuse, and the time for a blob to find a neighbour and coalesce with it is given by $a^{2} \sim Dt$ (with $D$ the diffusion constant). Thus in time $t = cR^3$ (with $c$ a proportionality constant) the mass of a blob doubles since we have just coalesced two neighbours.
Hence, the evolution of the characteristic blob mass obeys the scaling relation
$[t, M_{blob}] \rightarrow [t + cR^3, 2M_{blob}]$; thus we have $t+cR^3\sim kt$, with $k = \mathcal{O}(1)$, for all $t$, and hence $t \sim cR^3$ and so
$R\sim t^{1/3}$, in excellent agreement with the observed value (Figs.~\ref{fig:fourier}f and \ref{fig:sim}b). 
Repeating the argument for a diffusion constant varying as $D\sim 1/R$ implies that $R(t)\sim t^{1/4}$.

To conclude, the active motion of {\it Tubifex} worms leads to a phase-separation mechanism that is different from that of normal polymers, and that seems to be due only to the active nature of the living worms and to the immobilization of the worms at the inside of the entangled blobs; it may therefore be a generic phenomenon, also occurring in other types of polymeric active matter with entanglement interactions. We hope that the results presented here will stimulate further experimental and theoretical work in this direction.

\textit{Acknowledgement.} We thank the UvA's workshop for technical assistance, aquarium shop Holgen for the fresh batches of {\it Tubifex}, and Moslem Mbehjati for experimental assistance. A.D.\ acknowledges funding from the EU's Horizon 2020 research and innovation program under the Individual Marie Sk\l{}odowska-Curie fellowship grant agreement  798455.
contact details: a.deblais@uva.nl; s.woutersen@uva.nl;
;d.bonn@uva.nl


\begin{thebibliography}{36}%
\makeatletter
\providecommand \@ifxundefined [1]{%
 \@ifx{#1\undefined}
}%
\providecommand \@ifnum [1]{%
 \ifnum #1\expandafter \@firstoftwo
 \else \expandafter \@secondoftwo
 \fi
}%
\providecommand \@ifx [1]{%
 \ifx #1\expandafter \@firstoftwo
 \else \expandafter \@secondoftwo
 \fi
}%
\providecommand \natexlab [1]{#1}%
\providecommand \enquote  [1]{``#1''}%
\providecommand \bibnamefont  [1]{#1}%
\providecommand \bibfnamefont [1]{#1}%
\providecommand \citenamefont [1]{#1}%
\providecommand \href@noop [0]{\@secondoftwo}%
\providecommand \href [0]{\begingroup \@sanitize@url \@href}%
\providecommand \@href[1]{\@@startlink{#1}\@@href}%
\providecommand \@@href[1]{\endgroup#1\@@endlink}%
\providecommand \@sanitize@url [0]{\catcode `\\12\catcode `\$12\catcode
  `\&12\catcode `\#12\catcode `\^12\catcode `\_12\catcode `\%12\relax}%
\providecommand \@@startlink[1]{}%
\providecommand \@@endlink[0]{}%
\providecommand \url  [0]{\begingroup\@sanitize@url \@url }%
\providecommand \@url [1]{\endgroup\@href {#1}{\urlprefix }}%
\providecommand \urlprefix  [0]{URL }%
\providecommand \Eprint [0]{\href }%
\providecommand \doibase [0]{http://dx.doi.org/}%
\providecommand \selectlanguage [0]{\@gobble}%
\providecommand \bibinfo  [0]{\@secondoftwo}%
\providecommand \bibfield  [0]{\@secondoftwo}%
\providecommand \translation [1]{[#1]}%
\providecommand \BibitemOpen [0]{}%
\providecommand \bibitemStop [0]{}%
\providecommand \bibitemNoStop [0]{.\EOS\space}%
\providecommand \EOS [0]{\spacefactor3000\relax}%
\providecommand \BibitemShut  [1]{\csname bibitem#1\endcsname}%
\let\auto@bib@innerbib\@empty
\bibitem [{\citenamefont {Palacci}\ \emph {et~al.}(2013)\citenamefont
  {Palacci}, \citenamefont {Sacanna}, \citenamefont {Steinberg}, \citenamefont
  {Pine},\ and\ \citenamefont {Chaikin}}]{palacci13}%
  \BibitemOpen
  \bibfield  {author} {\bibinfo {author} {\bibfnamefont {J.}~\bibnamefont
  {Palacci}}, \bibinfo {author} {\bibfnamefont {S.}~\bibnamefont {Sacanna}},
  \bibinfo {author} {\bibfnamefont {A.~P.}\ \bibnamefont {Steinberg}}, \bibinfo
  {author} {\bibfnamefont {D.~J.}\ \bibnamefont {Pine}}, \ and\ \bibinfo
  {author} {\bibfnamefont {P.~M.}\ \bibnamefont {Chaikin}},\ }\href@noop {}
  {\bibfield  {journal} {\bibinfo  {journal} {Science}\ }\textbf {\bibinfo
  {volume} {339}},\ \bibinfo {pages} {936} (\bibinfo {year}
  {2013})}\BibitemShut {NoStop}%
\bibitem [{\citenamefont {Geyer}\ \emph {et~al.}(2019)\citenamefont {Geyer},
  \citenamefont {Martin}, \citenamefont {Tailleur},\ and\ \citenamefont
  {Bartolo}}]{Geyer2019}%
  \BibitemOpen
  \bibfield  {author} {\bibinfo {author} {\bibfnamefont {D.}~\bibnamefont
  {Geyer}}, \bibinfo {author} {\bibfnamefont {D.}~\bibnamefont {Martin}},
  \bibinfo {author} {\bibfnamefont {J.}~\bibnamefont {Tailleur}}, \ and\
  \bibinfo {author} {\bibfnamefont {D.}~\bibnamefont {Bartolo}},\ }\href@noop
  {} {\bibfield  {journal} {\bibinfo  {journal} {Phys. Rev. X}\ }\textbf
  {\bibinfo {volume} {9}},\ \bibinfo {pages} {031043} (\bibinfo {year}
  {2019})}\BibitemShut {NoStop}%
\bibitem [{\citenamefont {van~der Linden}\ \emph {et~al.}(2019)\citenamefont
  {van~der Linden}, \citenamefont {Alexander}, \citenamefont {Aarts},\ and\
  \citenamefont {Dauchot}}]{Linden2019}%
  \BibitemOpen
  \bibfield  {author} {\bibinfo {author} {\bibfnamefont {M.~N.}\ \bibnamefont
  {van~der Linden}}, \bibinfo {author} {\bibfnamefont {L.~C.}\ \bibnamefont
  {Alexander}}, \bibinfo {author} {\bibfnamefont {D.~G.}\ \bibnamefont
  {Aarts}}, \ and\ \bibinfo {author} {\bibfnamefont {O.}~\bibnamefont
  {Dauchot}},\ }\href@noop {} {\bibfield  {journal} {\bibinfo  {journal} {Phys.
  Rev. Lett.}\ }\textbf {\bibinfo {volume} {123}},\ \bibinfo {pages} {098001}
  (\bibinfo {year} {2019})}\BibitemShut {NoStop}%
\bibitem [{\citenamefont {Briand}\ and\ \citenamefont
  {Dauchot}(2016)}]{Briand2016}%
  \BibitemOpen
  \bibfield  {author} {\bibinfo {author} {\bibfnamefont {G.}~\bibnamefont
  {Briand}}\ and\ \bibinfo {author} {\bibfnamefont {O.}~\bibnamefont
  {Dauchot}},\ }\href@noop {} {\bibfield  {journal} {\bibinfo  {journal} {Phys.
  Rev. Lett.}\ }\textbf {\bibinfo {volume} {117}},\ \bibinfo {pages} {098004}
  (\bibinfo {year} {2016})}\BibitemShut {NoStop}%
\bibitem [{\citenamefont {Deblais}\ \emph {et~al.}(2018)\citenamefont
  {Deblais}, \citenamefont {Barois}, \citenamefont {Delville}, \citenamefont
  {Vaudaine}, \citenamefont {Lintuvuori}, \citenamefont {Boudet}, \citenamefont
  {Baret},\ and\ \citenamefont {Kellay}}]{Deblais2018}%
  \BibitemOpen
  \bibfield  {author} {\bibinfo {author} {\bibfnamefont {A.}~\bibnamefont
  {Deblais}}, \bibinfo {author} {\bibfnamefont {T.}~\bibnamefont {Barois}},
  \bibinfo {author} {\bibfnamefont {P.}~\bibnamefont {Delville}}, \bibinfo
  {author} {\bibfnamefont {R.}~\bibnamefont {Vaudaine}}, \bibinfo {author}
  {\bibfnamefont {J.~S.}\ \bibnamefont {Lintuvuori}}, \bibinfo {author}
  {\bibfnamefont {J.}~\bibnamefont {Boudet}}, \bibinfo {author} {\bibfnamefont
  {J.}~\bibnamefont {Baret}}, \ and\ \bibinfo {author} {\bibfnamefont
  {H.}~\bibnamefont {Kellay}},\ }\href@noop {} {\bibfield  {journal} {\bibinfo
  {journal} {Phys. Rev. Lett.}\ }\textbf {\bibinfo {volume} {120}},\ \bibinfo
  {pages} {188002} (\bibinfo {year} {2018})}\BibitemShut {NoStop}%
\bibitem [{\citenamefont {Schwarz-Linek}\ \emph {et~al.}(2012)\citenamefont
  {Schwarz-Linek}, \citenamefont {Valeriani}, \citenamefont {Cacciuto},
  \citenamefont {Cates}, \citenamefont {Marenduzzo}, \citenamefont {Morozov},\
  and\ \citenamefont {Poon}}]{Schwarz-Lineka2012}%
  \BibitemOpen
  \bibfield  {author} {\bibinfo {author} {\bibfnamefont {J.}~\bibnamefont
  {Schwarz-Linek}}, \bibinfo {author} {\bibfnamefont {C.}~\bibnamefont
  {Valeriani}}, \bibinfo {author} {\bibfnamefont {A.}~\bibnamefont {Cacciuto}},
  \bibinfo {author} {\bibfnamefont {M.~E.}\ \bibnamefont {Cates}}, \bibinfo
  {author} {\bibfnamefont {D.}~\bibnamefont {Marenduzzo}}, \bibinfo {author}
  {\bibfnamefont {A.~N.}\ \bibnamefont {Morozov}}, \ and\ \bibinfo {author}
  {\bibfnamefont {W.~C.~K.}\ \bibnamefont {Poon}},\ }\href@noop {} {\bibfield
  {journal} {\bibinfo  {journal} {Proc Natl Acad Sci USA}\ }\textbf {\bibinfo
  {volume} {109}},\ \bibinfo {pages} {4052} (\bibinfo {year}
  {2012})}\BibitemShut {NoStop}%
\bibitem [{\citenamefont {Duclos}\ \emph {et~al.}(2018)\citenamefont {Duclos},
  \citenamefont {Blanch-Mercader}, \citenamefont {Yashunsky}, \citenamefont
  {Salbreux}, \citenamefont {Joanny}, \citenamefont {Prost},\ and\
  \citenamefont {Silberzan}}]{Duclos2018}%
  \BibitemOpen
  \bibfield  {author} {\bibinfo {author} {\bibfnamefont {G.}~\bibnamefont
  {Duclos}}, \bibinfo {author} {\bibfnamefont {C.}~\bibnamefont
  {Blanch-Mercader}}, \bibinfo {author} {\bibfnamefont {V.}~\bibnamefont
  {Yashunsky}}, \bibinfo {author} {\bibfnamefont {G.}~\bibnamefont {Salbreux}},
  \bibinfo {author} {\bibfnamefont {J.-F.}\ \bibnamefont {Joanny}}, \bibinfo
  {author} {\bibfnamefont {J.}~\bibnamefont {Prost}}, \ and\ \bibinfo {author}
  {\bibfnamefont {P.}~\bibnamefont {Silberzan}},\ }\href@noop {} {\bibfield
  {journal} {\bibinfo  {journal} {Nature Physics}\ }\textbf {\bibinfo {volume}
  {14}},\ \bibinfo {pages} {728} (\bibinfo {year} {2018})}\BibitemShut
  {NoStop}%
\bibitem [{\citenamefont {Tennenbaum}\ \emph {et~al.}(2015)\citenamefont
  {Tennenbaum}, \citenamefont {Liu}, \citenamefont {Hu},\ and\ \citenamefont
  {Fernandez-Nieves}}]{Tennenbaum2015}%
  \BibitemOpen
  \bibfield  {author} {\bibinfo {author} {\bibfnamefont {M.}~\bibnamefont
  {Tennenbaum}}, \bibinfo {author} {\bibfnamefont {Z.}~\bibnamefont {Liu}},
  \bibinfo {author} {\bibfnamefont {D.}~\bibnamefont {Hu}}, \ and\ \bibinfo
  {author} {\bibfnamefont {A.}~\bibnamefont {Fernandez-Nieves}},\ }\href@noop
  {} {\bibfield  {journal} {\bibinfo  {journal} {Nat. Mat.}\ }\textbf {\bibinfo
  {volume} {15}},\ \bibinfo {pages} {54} (\bibinfo {year} {2015})}\BibitemShut
  {NoStop}%
\bibitem [{\citenamefont {Sugi}\ \emph {et~al.}(2019)\citenamefont {Sugi},
  \citenamefont {Ito}, \citenamefont {Nishimura},\ and\ \citenamefont
  {Nagai}}]{sugi19}%
  \BibitemOpen
  \bibfield  {author} {\bibinfo {author} {\bibfnamefont {T.}~\bibnamefont
  {Sugi}}, \bibinfo {author} {\bibfnamefont {H.}~\bibnamefont {Ito}}, \bibinfo
  {author} {\bibfnamefont {M.}~\bibnamefont {Nishimura}}, \ and\ \bibinfo
  {author} {\bibfnamefont {K.~H.}\ \bibnamefont {Nagai}},\ }\href@noop {}
  {\bibfield  {journal} {\bibinfo  {journal} {Nat. Comm.}\ }\textbf {\bibinfo
  {volume} {10}},\ \bibinfo {pages} {683} (\bibinfo {year} {2019})}\BibitemShut
  {NoStop}%
\bibitem [{\citenamefont {Bain}\ and\ \citenamefont
  {Bartolo}(2019)}]{Bain2019}%
  \BibitemOpen
  \bibfield  {author} {\bibinfo {author} {\bibfnamefont {N.}~\bibnamefont
  {Bain}}\ and\ \bibinfo {author} {\bibfnamefont {D.}~\bibnamefont {Bartolo}},\
  }\href@noop {} {\bibfield  {journal} {\bibinfo  {journal} {Science}\ }\textbf
  {\bibinfo {volume} {363}},\ \bibinfo {pages} {46} (\bibinfo {year}
  {2019})}\BibitemShut {NoStop}%
\bibitem [{\citenamefont {Liverpool}\ \emph {et~al.}(2000)\citenamefont
  {Liverpool}, \citenamefont {Maggs},\ and\ \citenamefont
  {Ajdari}}]{liverpool00}%
  \BibitemOpen
  \bibfield  {author} {\bibinfo {author} {\bibfnamefont {T.~B.}\ \bibnamefont
  {Liverpool}}, \bibinfo {author} {\bibfnamefont {A.~C.}\ \bibnamefont
  {Maggs}}, \ and\ \bibinfo {author} {\bibfnamefont {A.}~\bibnamefont
  {Ajdari}},\ }\href@noop {} {\bibfield  {journal} {\bibinfo  {journal} {Phys.
  Rev. Lett.}\ }\textbf {\bibinfo {volume} {86}},\ \bibinfo {pages} {4171}
  (\bibinfo {year} {2000})}\BibitemShut {NoStop}%
\bibitem [{\citenamefont {Speck}\ \emph {et~al.}(2014)\citenamefont {Speck},
  \citenamefont {Bialk\'{e}}, \citenamefont {Menzel},\ and\ \citenamefont
  {L\"{o}wen}}]{speck14}%
  \BibitemOpen
  \bibfield  {author} {\bibinfo {author} {\bibfnamefont {T.}~\bibnamefont
  {Speck}}, \bibinfo {author} {\bibfnamefont {J.}~\bibnamefont {Bialk\'{e}}},
  \bibinfo {author} {\bibfnamefont {A.~M.}\ \bibnamefont {Menzel}}, \ and\
  \bibinfo {author} {\bibfnamefont {H.}~\bibnamefont {L\"{o}wen}},\ }\href@noop
  {} {\bibfield  {journal} {\bibinfo  {journal} {Phys. Rev. Lett.}\ }\textbf
  {\bibinfo {volume} {112}},\ \bibinfo {pages} {218304} (\bibinfo {year}
  {2014})}\BibitemShut {NoStop}%
\bibitem [{\citenamefont {Bechinger}\ \emph {et~al.}(2012)\citenamefont
  {Bechinger}, \citenamefont {Leonardo}, \citenamefont {Cacciutob},\ and\
  \citenamefont {Cates}}]{Bechinger2012}%
  \BibitemOpen
  \bibfield  {author} {\bibinfo {author} {\bibfnamefont {C.}~\bibnamefont
  {Bechinger}}, \bibinfo {author} {\bibfnamefont {R.~D.}\ \bibnamefont
  {Leonardo}}, \bibinfo {author} {\bibfnamefont {A.}~\bibnamefont {Cacciutob}},
  \ and\ \bibinfo {author} {\bibfnamefont {M.~E.}\ \bibnamefont {Cates}},\
  }\href@noop {} {\bibfield  {journal} {\bibinfo  {journal} {Rev. Mod. Phys.}\
  }\textbf {\bibinfo {volume} {88}},\ \bibinfo {pages} {045006} (\bibinfo
  {year} {2012})}\BibitemShut {NoStop}%
\bibitem [{\citenamefont {Dobry}\ and\ \citenamefont
  {Boyer-Kawenoki}(1947)}]{Dobry1947}%
  \BibitemOpen
  \bibfield  {author} {\bibinfo {author} {\bibfnamefont {A.}~\bibnamefont
  {Dobry}}\ and\ \bibinfo {author} {\bibfnamefont {F.}~\bibnamefont
  {Boyer-Kawenoki}},\ }\href@noop {} {\bibfield  {journal} {\bibinfo  {journal}
  {Journal of Polymer Science}\ }\textbf {\bibinfo {volume} {2}},\ \bibinfo
  {pages} {90} (\bibinfo {year} {1947})}\BibitemShut {NoStop}%
\bibitem [{\citenamefont {Tanaka}\ and\ \citenamefont
  {Nishikawa}(2005)}]{Tanaka2005}%
  \BibitemOpen
  \bibfield  {author} {\bibinfo {author} {\bibfnamefont {H.}~\bibnamefont
  {Tanaka}}\ and\ \bibinfo {author} {\bibfnamefont {Y.}~\bibnamefont
  {Nishikawa}},\ }\href@noop {} {\bibfield  {journal} {\bibinfo  {journal}
  {Phys. Rev. Lett.}\ }\textbf {\bibinfo {volume} {95}},\ \bibinfo {pages}
  {078103} (\bibinfo {year} {2005})}\BibitemShut {NoStop}%
\bibitem [{\citenamefont {Olson}\ and\ \citenamefont {Roth}(1988)}]{Olson1988}%
  \BibitemOpen
  \bibfield  {author} {\bibinfo {author} {\bibfnamefont {G.}~\bibnamefont
  {Olson}}\ and\ \bibinfo {author} {\bibfnamefont {J.}~\bibnamefont {Roth}},\
  }\href@noop {} {\bibfield  {journal} {\bibinfo  {journal} {Materials Science
  Reports}\ }\textbf {\bibinfo {volume} {3}},\ \bibinfo {pages} {1} (\bibinfo
  {year} {1988})}\BibitemShut {NoStop}%
\bibitem [{\citenamefont {Voorhees}(1992)}]{Voorhees1992}%
  \BibitemOpen
  \bibfield  {author} {\bibinfo {author} {\bibfnamefont {P.~W.}\ \bibnamefont
  {Voorhees}},\ }\href@noop {} {\bibfield  {journal} {\bibinfo  {journal}
  {Annu. Rev. Maler. Sci.}\ }\textbf {\bibinfo {volume} {22}},\ \bibinfo
  {pages} {197} (\bibinfo {year} {1992})}\BibitemShut {NoStop}%
\bibitem [{\citenamefont {A.~G.~Thompson}\ and\ \citenamefont
  {Blythe}(2011)}]{Thompson2011}%
  \BibitemOpen
  \bibfield  {author} {\bibinfo {author} {\bibfnamefont {M.~E.~C.}\
  \bibnamefont {A.~G.~Thompson}, \bibfnamefont {J.~Tailleur}}\ and\ \bibinfo
  {author} {\bibfnamefont {R.~A.}\ \bibnamefont {Blythe}},\ }\href@noop {}
  {\bibfield  {journal} {\bibinfo  {journal} {J. Stat. Mech.}\ }\textbf
  {\bibinfo {volume} {2}},\ \bibinfo {pages} {02029} (\bibinfo {year}
  {2011})}\BibitemShut {NoStop}%
\bibitem [{\citenamefont {Cates}\ and\ \citenamefont
  {Tailleur}(2015)}]{Cates2015}%
  \BibitemOpen
  \bibfield  {author} {\bibinfo {author} {\bibfnamefont {M.~E.}\ \bibnamefont
  {Cates}}\ and\ \bibinfo {author} {\bibfnamefont {J.}~\bibnamefont
  {Tailleur}},\ }\href@noop {} {\bibfield  {journal} {\bibinfo  {journal}
  {Annu. Rev. Condens. Matter Phys.}\ }\textbf {\bibinfo {volume} {6}},\
  \bibinfo {pages} {219} (\bibinfo {year} {2015})}\BibitemShut {NoStop}%
\bibitem [{\citenamefont {Gonnella}\ \emph {et~al.}(2015)\citenamefont
  {Gonnella}, \citenamefont {Marenduzzo}, \citenamefont {Suma},\ and\
  \citenamefont {Tiribocchi}}]{gonnella15}%
  \BibitemOpen
  \bibfield  {author} {\bibinfo {author} {\bibfnamefont {G.}~\bibnamefont
  {Gonnella}}, \bibinfo {author} {\bibfnamefont {D.}~\bibnamefont
  {Marenduzzo}}, \bibinfo {author} {\bibfnamefont {A.}~\bibnamefont {Suma}}, \
  and\ \bibinfo {author} {\bibfnamefont {A.}~\bibnamefont {Tiribocchi}},\
  }\href@noop {} {\bibfield  {journal} {\bibinfo  {journal} {C. R. Phys.}\
  }\textbf {\bibinfo {volume} {16}},\ \bibinfo {pages} {316} (\bibinfo {year}
  {2015})}\BibitemShut {NoStop}%
\bibitem [{sup()}]{supmat}%
  \BibitemOpen
  \href@noop {} {}\bibinfo {note} {See Supplemental Material for experimental
  details and additional results. It includes refs
  \cite{Walker1970,deGennes2013,sugi19,solon}.}\BibitemShut {Stop}%
\bibitem [{\citenamefont {Winkler}\ \emph {et~al.}(2017)\citenamefont
  {Winkler}, \citenamefont {Elgeti},\ and\ \citenamefont
  {Gompper}}]{winkler17}%
  \BibitemOpen
  \bibfield  {author} {\bibinfo {author} {\bibfnamefont {R.~G.}\ \bibnamefont
  {Winkler}}, \bibinfo {author} {\bibfnamefont {J.}~\bibnamefont {Elgeti}}, \
  and\ \bibinfo {author} {\bibfnamefont {G.}~\bibnamefont {Gompper}},\
  }\href@noop {} {\bibfield  {journal} {\bibinfo  {journal} {J. Phys. Soc.
  Jpn.}\ }\textbf {\bibinfo {volume} {86}},\ \bibinfo {pages} {101014}
  (\bibinfo {year} {2017})}\BibitemShut {NoStop}%
\bibitem [{\citenamefont {Mart\'{i}n-G\'{o}́mez}\ \emph
  {et~al.}(2018)\citenamefont {Mart\'{i}n-G\'{o}́mez}, \citenamefont
  {Gompper},\ and\ \citenamefont {Winkler}}]{martin18}%
  \BibitemOpen
  \bibfield  {author} {\bibinfo {author} {\bibfnamefont {A.}~\bibnamefont
  {Mart\'{i}n-G\'{o}́mez}}, \bibinfo {author} {\bibfnamefont {G.}~\bibnamefont
  {Gompper}}, \ and\ \bibinfo {author} {\bibfnamefont {R.~G.}\ \bibnamefont
  {Winkler}},\ }\href@noop {} {\bibfield  {journal} {\bibinfo  {journal}
  {Polymers}\ }\textbf {\bibinfo {volume} {2018}},\ \bibinfo {pages} {837}
  (\bibinfo {year} {2018})}\BibitemShut {NoStop}%
\bibitem [{\citenamefont {Bianco}\ \emph {et~al.}(2018)\citenamefont {Bianco},
  \citenamefont {Locatelli},\ and\ \citenamefont {Malgaretti}}]{Bianco2018}%
  \BibitemOpen
  \bibfield  {author} {\bibinfo {author} {\bibfnamefont {V.}~\bibnamefont
  {Bianco}}, \bibinfo {author} {\bibfnamefont {E.}~\bibnamefont {Locatelli}}, \
  and\ \bibinfo {author} {\bibfnamefont {P.}~\bibnamefont {Malgaretti}},\
  }\href@noop {} {\bibfield  {journal} {\bibinfo  {journal} {Phys. Rev. Lett.}\
  }\textbf {\bibinfo {volume} {121}},\ \bibinfo {pages} {217802} (\bibinfo
  {year} {2018})}\BibitemShut {NoStop}%
\bibitem [{\citenamefont {Mart\'{i}n-G\'{o}́mez}\ \emph
  {et~al.}(2019)\citenamefont {Mart\'{i}n-G\'{o}́mez}, \citenamefont
  {Eisenstecken}, \citenamefont {Gompper},\ and\ \citenamefont
  {Winkler}}]{martin19}%
  \BibitemOpen
  \bibfield  {author} {\bibinfo {author} {\bibfnamefont {A.}~\bibnamefont
  {Mart\'{i}n-G\'{o}́mez}}, \bibinfo {author} {\bibfnamefont {T.}~\bibnamefont
  {Eisenstecken}}, \bibinfo {author} {\bibfnamefont {G.}~\bibnamefont
  {Gompper}}, \ and\ \bibinfo {author} {\bibfnamefont {R.~G.}\ \bibnamefont
  {Winkler}},\ }\href@noop {} {\bibfield  {journal} {\bibinfo  {journal} {Soft
  Matter}\ }\textbf {\bibinfo {volume} {15}},\ \bibinfo {pages} {3957}
  (\bibinfo {year} {2019})}\BibitemShut {NoStop}%
\bibitem [{\citenamefont {Mokhtari}\ and\ \citenamefont
  {Zippelius}(2019)}]{Mokthari2019}%
  \BibitemOpen
  \bibfield  {author} {\bibinfo {author} {\bibfnamefont {Z.}~\bibnamefont
  {Mokhtari}}\ and\ \bibinfo {author} {\bibfnamefont {A.}~\bibnamefont
  {Zippelius}},\ }\href@noop {} {\bibfield  {journal} {\bibinfo  {journal}
  {Phys. Rev. Lett.}\ }\textbf {\bibinfo {volume} {123}},\ \bibinfo {pages}
  {028001} (\bibinfo {year} {2019})}\BibitemShut {NoStop}%
\bibitem [{\citenamefont {Deblais}\ \emph {et~al.}(2019)\citenamefont
  {Deblais}, \citenamefont {Woutersen},\ and\ \citenamefont
  {Bonn}}]{deblais_preprint_2019}%
  \BibitemOpen
  \bibfield  {author} {\bibinfo {author} {\bibfnamefont {A.}~\bibnamefont
  {Deblais}}, \bibinfo {author} {\bibfnamefont {S.}~\bibnamefont {Woutersen}},
  \ and\ \bibinfo {author} {\bibfnamefont {D.}~\bibnamefont {Bonn}},\
  }\href@noop {} {\enquote {\bibinfo {title} {Rheology of active-polymer-like
  {T}.~{Tubifex} worms},}\ } (\bibinfo {year} {2019}),\ \Eprint
  {http://arxiv.org/abs/arXiv:1910.09612} {arXiv:1910.09612} \BibitemShut
  {NoStop}%
\bibitem [{\citenamefont {Walker}(1971)}]{walker1971}%
  \BibitemOpen
  \bibfield  {author} {\bibinfo {author} {\bibfnamefont {J.~G.}\ \bibnamefont
  {Walker}},\ }\href@noop {} {\bibfield  {journal} {\bibinfo  {journal} {Biol.
  Bull.}\ }\textbf {\bibinfo {volume} {138}},\ \bibinfo {pages} {235} (\bibinfo
  {year} {1971})}\BibitemShut {NoStop}%
\bibitem [{\citenamefont {Brinkhurst}(1963)}]{brinkhurst63}%
  \BibitemOpen
  \bibinfo {editor} {\bibfnamefont {R.~O.}\ \bibnamefont {Brinkhurst}},\ ed.,\
  \href@noop {} {\emph {\bibinfo {title} {Taxonomical Studies on the
  Tubificidae}}}\ (\bibinfo  {publisher} {Akadamie Verlag},\ \bibinfo {address}
  {Berlin},\ \bibinfo {year} {1963})\BibitemShut {NoStop}%
\bibitem [{\citenamefont {Jones}(2002)}]{jonesbook}%
  \BibitemOpen
  \bibfield  {author} {\bibinfo {author} {\bibfnamefont {R.~A.~L.}\
  \bibnamefont {Jones}},\ }\href@noop {} {\emph {\bibinfo {title} {Soft
  Condensed Matter}}}\ (\bibinfo  {publisher} {Oxford University Press},\
  \bibinfo {address} {Oxford},\ \bibinfo {year} {2002})\BibitemShut {NoStop}%
\bibitem [{\citenamefont {Family}\ and\ \citenamefont
  {Meakin}(1988)}]{family88}%
  \BibitemOpen
  \bibfield  {author} {\bibinfo {author} {\bibfnamefont {F.}~\bibnamefont
  {Family}}\ and\ \bibinfo {author} {\bibfnamefont {P.}~\bibnamefont
  {Meakin}},\ }\href@noop {} {\bibfield  {journal} {\bibinfo  {journal} {Phys.
  Rev. Lett.}\ }\textbf {\bibinfo {volume} {61}},\ \bibinfo {pages} {428}
  (\bibinfo {year} {1988})}\BibitemShut {NoStop}%
\bibitem [{\citenamefont {Steyer}\ \emph {et~al.}(1991)\citenamefont {Steyer},
  \citenamefont {Guenon}, \citenamefont {Beyens},\ and\ \citenamefont
  {Knobler}}]{steyer91}%
  \BibitemOpen
  \bibfield  {author} {\bibinfo {author} {\bibfnamefont {A.}~\bibnamefont
  {Steyer}}, \bibinfo {author} {\bibfnamefont {P.}~\bibnamefont {Guenon}},
  \bibinfo {author} {\bibfnamefont {D.}~\bibnamefont {Beyens}}, \ and\ \bibinfo
  {author} {\bibfnamefont {C.~M.}\ \bibnamefont {Knobler}},\ }\href@noop {}
  {\bibfield  {journal} {\bibinfo  {journal} {Pys. Rev. A}\ }\textbf {\bibinfo
  {volume} {44}},\ \bibinfo {pages} {8271} (\bibinfo {year}
  {1991})}\BibitemShut {NoStop}%
\bibitem [{\citenamefont {Meakin}(1992)}]{meakin92}%
  \BibitemOpen
  \bibfield  {author} {\bibinfo {author} {\bibfnamefont {P.}~\bibnamefont
  {Meakin}},\ }\href@noop {} {\bibfield  {journal} {\bibinfo  {journal} {Rep.
  Prog. Phys.}\ }\textbf {\bibinfo {volume} {55}},\ \bibinfo {pages} {157}
  (\bibinfo {year} {1992})}\BibitemShut {NoStop}%
\bibitem [{\citenamefont {De~Gennes}\ \emph {et~al.}(2013)\citenamefont
  {De~Gennes}, \citenamefont {Brochard-Wyart},\ and\ \citenamefont
  {Qu{\'e}r{\'e}}}]{deGennes2013}%
  \BibitemOpen
  \bibfield  {author} {\bibinfo {author} {\bibfnamefont {P.-G.}\ \bibnamefont
  {De~Gennes}}, \bibinfo {author} {\bibfnamefont {F.}~\bibnamefont
  {Brochard-Wyart}}, \ and\ \bibinfo {author} {\bibfnamefont {D.}~\bibnamefont
  {Qu{\'e}r{\'e}}},\ }\href@noop {} {\emph {\bibinfo {title} {Capillarity and
  wetting phenomena: drops, bubbles, pearls, waves}}}\ (\bibinfo  {publisher}
  {Springer Science \& Business Media},\ \bibinfo {year} {2013})\BibitemShut
  {NoStop}%
\bibitem [{\citenamefont {Walker}(1970)}]{Walker1970}%
  \BibitemOpen
  \bibfield  {author} {\bibinfo {author} {\bibfnamefont {J.~G.}\ \bibnamefont
  {Walker}},\ }\href@noop {} {\bibfield  {journal} {\bibinfo  {journal} {Biol.
  Bull.}\ ,\ \bibinfo {pages} {235}} (\bibinfo {year} {1970})}\BibitemShut
  {NoStop}%
\bibitem [{\citenamefont {Solon}\ \emph {et~al.}(2015)\citenamefont {Solon},
  \citenamefont {Fily}, \citenamefont {Baskaran}, \citenamefont {Cates},
  \citenamefont {Kafri}, \citenamefont {Kardar},\ and\ \citenamefont
  {Tailleur}}]{solon}%
  \BibitemOpen
  \bibfield  {author} {\bibinfo {author} {\bibfnamefont {A.~P.}\ \bibnamefont
  {Solon}}, \bibinfo {author} {\bibfnamefont {Y.}~\bibnamefont {Fily}},
  \bibinfo {author} {\bibfnamefont {A.}~\bibnamefont {Baskaran}}, \bibinfo
  {author} {\bibfnamefont {M.~E.}\ \bibnamefont {Cates}}, \bibinfo {author}
  {\bibfnamefont {Y.}~\bibnamefont {Kafri}}, \bibinfo {author} {\bibfnamefont
  {M.}~\bibnamefont {Kardar}}, \ and\ \bibinfo {author} {\bibfnamefont
  {J.}~\bibnamefont {Tailleur}},\ }\href@noop {} {\bibfield  {journal}
  {\bibinfo  {journal} {Nature Physics}\ }\textbf {\bibinfo {volume} {11}},\
  \bibinfo {pages} {673} (\bibinfo {year} {2015})}\BibitemShut {NoStop}%
\end{thebibliography}

%

\end{document}